\documentclass[reprint, superscriptaddress, aps, pre]{revtex4-1}

\usepackage[T1]{fontenc}
\usepackage{graphicx}
\usepackage{xfrac}

\usepackage[
  unicode,
  pdfusetitle,    % title, authors and date as pdf attributes
  pdfcreator={},
  pdfproducer={},
]{hyperref}

\usepackage{mathtools}

\let\theta\vartheta
\newcommand*{\dd}{\mathrm{d}}
\newcommand*{\pp}{\mathrm{p}}
\newcommand*{\bu}{\textbf{u}}
\newcommand*{\bq}{\textbf{q}}
\newcommand*{\br}{\textbf{r}}
\newcommand*{\bnod}{\textbf{0}}
\newcommand*{\bM}{\textbf{M}}
 
\newcommand*{\etal}{\textit{et al.}}
\newcommand*{\kbt}{k_\text{B}T}

\usepackage{xcolor}
\colorlet{myred}{red!80!black}

\begin{document}

%%%%%%%%% TITLE SECTION %%%%%%%%%%

\title{Geometric percolation of hard nanorods: the interplay of spontaneous and externally induced uniaxial particle alignment}
\author{Shari P. Finner}
\email{s.p.finner@tue.nl}
\affiliation{Department of Applied Physics, Eindhoven University of Technology, P.O. Box 513,
3500 MB Eindhoven, The Netherlands}
\author{Ilian Pihlajamaa}
\affiliation{Department of Applied Physics, Eindhoven University of Technology, P.O. Box 513,
3500 MB Eindhoven, The Netherlands}
\author{Paul van der Schoot}
\affiliation{Department of Applied Physics, Eindhoven University of Technology, P.O. Box 513,
3500 MB Eindhoven, The Netherlands}
\affiliation{Institute for Theoretical Physics, Utrecht University, Princetonplein 5, 3584 CC Utrecht, The Netherlands}

\begin{abstract}
	We present a numerical study on geometric percolation in liquid dispersions of hard slender colloidal particles subjected to an external orienting field. 
	In the  formulation and liquid-state processing of nanocomposite materials, the alignment of particles by external fields such as electric, magnetic or flow fields is practically inevitable, and often works against the emergence of large nanoparticle networks.
	Using continuum percolation theory in conjunction with Onsager theory, we investigate how the interplay between \textit{externally} induced alignment and the \textit{spontaneous} symmetry breaking of the uniaxial nematic phase affects cluster formation within nanoparticle dispersions.
	It is known that the enhancement of particle alignment by means of a density increase or an external field may result in the breakdown of an already percolating network.
	As a result, percolation can be limited to a small region of the phase diagram only.
	Here, we demonstrate that the existence and shape of such a ``percolation island'' in the phase diagram crucially depends on the connectivity length -- a critical distance defining direct connections between neighbouring particles. 
	Deformations of this percolation island can lead to peculiar re-entrance effects, in which a system-spanning network forms and breaks down multiple times with increasing particle density.
\end{abstract}

\maketitle

%%%%%%%%%%%%%%%%%%%%%%%%%%%%%%%%%%%%%%%%%%%%%%%%%%%%%%

	\section{Introduction}
		The addition of nanoparticles to polymeric host materials is a powerful tool in the production and design of high-performance materials, as it can strongly enhance the mechanical, thermal or electrical properties of the host medium.~\cite{DresselhausCNTbook, KoningCNTbook}
		For a drastic enhancement of material properties, the nanoparticles typically need to be -- in some sense -- connected and form a material-spanning network.
		This happens above a critical particle loading called the percolation threshold.~\cite{Torquatobook}
		Adding a critical amount of conductive particles like carbon nanotubes, graphene, or silver nanowires to common engineering plastics, for instance, can radically increase the electrical and thermal conductivity of the resulting nanocomposite material. \cite{Maiti2014, KoningCNTbook}
		For the rational design of nanocomposites with the desired properties, a topic of broad relevance in the areas of optoelectronics, photovoltaics, and electromagnetic interference shielding, it is essential to understand and be able to control network formation in nanoparticle dispersions.~\cite{Moniruzzaman2006, Mutiso2015, Grossiord2008, Kim2002, Shah2015,Thomassin2013}
		
		Nanoparticles in liquid dispersions do not actually need to make physical contact for a conducting network to emerge, as effective charge transport can take place via quantum mechanical tunneling.~\cite{Grossiord2006, Shklovskii2006, Ambrosetti2010}
		While, according to standard quantum mechanics, the tunneling probability is an exponentially decaying function of the distance, assuming a hard cut-off distance yields the same results for the percolation threshold.~\cite{Otten2009, Otten2011}
		Therefore, we consider \textit{geometric percolation}, in which two particles count as connected if their surface-to-surface distance is lower than a critical connectivity range, which corresponds to the average tunneling distance of charge carriers through the polymeric host matrix.
		
		Next to the connectivity range, there are many factors that influence the formation of transient networks in nanoparticle dispersions.
		A vast body of literature has shown that the particle anisometry,~\cite{Schilling2015, Celzard1996, Finner2018, Balberg1984, Otten2011} size polydispersity,~\cite{Balberg1984, Balberg1986, Otten2009, Otten2011, Chatterjee2010, Nigro2013, Chatterjee2014, Meyer2015, Kale2015, Ambrosetti2010, Sapkota2017} the presence of nonconducting particles,~\cite{Otten2011} inter-particle interactions \cite{Kyrylyuk2008, Vigolo2005, Schilling2007} and external fields \cite{Otten2012, Finner2018} can have tremendous effects on the percolation threshold, while the influences of flexibility and the precise particle shape seem to be rather subtle. \cite{Kyrylyuk2008, Kwon2016, Drwenski2017}
		In this article we focus on the fact that many types of conductive nanoparticle, like carbon nanotubes or graphene, have a strongly anisotropic shape, which strongly affects their interactions and the way they react to external orienting fields.
		
		A common engineering goal is to keep the percolation threshold as low as possible in order to preserve other desired properties of the host material, like its optical transparency.~\cite{Mutiso2015, Grossiord2008, Ackermann2016}
		Strongly anisotropic nanofillers are particularly suited for these applications, as the (equilibrium) percolation threshold scales inversely with the nanoparticle aspect ratio, at least if the particles are isotropically oriented.~\cite{Garboczi1995, Balberg1986, Bug1985, Bug1986, Otten2009, Otten2011, Maiti2014, Grimaldi2017, Drwenski2017}
		The low percolation threshold of strongly anisometric particles can be attributed to their relatively large contact volume, \textit{i.e.}, the volume one particle can physically be in so that it is connected to a second, fixed test particle.~\cite{Balberg1984, Celzard1996}
		
		In real applications, however, anisometric particles are very often not isotropically oriented.
		Due to a competition between orientational and translational entropy, they can exhibit a transition from the isotropic phase to liquid crystalline phases with increasing particle density.~\cite{Onsager1949, Odijk1986review, Vroege1992, Singh2000review, Song2005}
		At  concentrations very close to the expected percolation threshold, slender rod-like particles and high aspect-ratio platelets form the uniaxial nematic liquid crystal phase and align along a common axis called the nematic director.
		For sufficiently small connectivity ranges, the percolation transition can even be pre-empted by the isotropic-nematic phase transition. \cite{Poulin2015, Atashpendar2018}
		
		Next to their liquid-crystalline behaviour, anisometric particles may get aligned during the processing of the nanocomposite in its liquid state.
		Depending on the application, alignment may actually be desired and intentionally induced.
		Applying electric or magnetic fields,~\cite{Bubke1997, Yamamoto1998, Chen2001, Martin2005, Brown2007, Chapkin2018, Zhu2009, Fujiwara2001, Zaric2004} or using a thermotropic liquid-crystalline host medium,~\cite{Lagerwall2008review, Lagerwall2016book, PopaNita2008, Dolgov2010book} for instance, can result in (tunable) anisotropic mechanical or transport properties of the nanocomposite. \cite{Chatterjee2018, Atashpendar2018, Zamora2011, Carmona1987, Wang2008}
		Anisometric filler particles may also align due to flow fields,~\cite{Wang2008, Du2005, Aguilar2016, Wescott2007, Lettinga2004} which arise during processing steps like extrusion, drawing, spin coating or slot die coating.
		 If the liquid is not allowed to relax before solidification, the aligned structure is then frozen in in the final composite. \cite{Grossiord2006, Skipa2010, Moradi2015}
		An often undesired effect of particle alignment is that it typically raises the percolation threshold by decreasing the contact volume of particle pairs and increasing the average surface-to-surface distance. \cite{Balberg1984, Chatterjee2014, Otten2012, Du2005, Grossiord2006, Deng2009, Rahatekar2005, White2009, Kale2016, Kumar2016}

		In a recent study, we have theoretically shown that percolation becomes very unusual if nanoparticles align by transitioning into the uniaxial nematic phase. \cite{Finner2019PRL}
		We found re-entrance percolation, which means that a percolating network can first be formed, and then be destroyed again with increasing particle concentration, at least if the connectivity range is sufficiently small.
		Here, we again combine Onsager theory with Continuum Percolation Theory, and additionally investigate the effect of an external orienting field.
		Our goal is to elucidate how the complex interplay between spontaneous nematic alignment and field-induced external alignment affects cluster formation and percolation in nanocomposites.
		
		The effect of combining hard-core particle interactions and an external alignment field has, in fact, already been studied within the same theoretical framework.~\cite{Otten2012}
		Using an analytical expansion of the two-particle contact volume in Legendre polynomials, Otten \etal~found interesting re-entrance percolation within the weakly ordered paranematic phase, which limits the occurrence of percolating networks to a small, contained ``island'' in the phase diagram that is bounded from above by the paranematic-nematic transition.
		However, the  findings in Ref.~\cite{Otten2012} partially disagree with our recent results in the uniaxial nematic phase \cite{Finner2019PRL} and raise a few questions that we intend to address in this article.
		
		With our numerical treatment of Onsager- and Continuum Percolation Theory, we show here that our former understanding of the interplay between external and intrinsic alignment was incomplete.
		For particles in the Onsager limit of infinite aspect ratios, we do observe the ``percolation island'' predicted by Ref.~\cite{Otten2012} for small values of the connectivity range, but find that re-entrance typically does not occur within the paranematic phase.
		Instead, percolation may be lost across the paranematic-nematic transition, or at even higher packing fractions inside the nematic phase.
		If the connectivity range is increased, the percolation region changes drastically in shape and ultimately opens up to an unbounded domain.
		Deformations of the percolation region can give rise to multiple re-entrance effects, in which percolation can be gained and lost twice upon the addition of particles.
		For particles of finite aspect ratio, we find that a percolating network can no longer be destroyed at high particle concentrations, consistent with our earlier prediction.~\cite{Finner2019PRE}
		
		To set up the theoretical framework, we first outline our model and describe Onsager theory for the distribution of particle orientations in Section \ref{sec:Onstheory} of this article.
		Section \ref{sec:percTheory} focuses on the description of cluster formation using Continuum Percolation Theory.
		In Section \ref{sec:numerics}, we briefly describe our numerical methods, before turning to our results for the percolation threshold in Section \ref{sec:percresults}.
		The physical cluster dimensions in terms of the correlation lengths are calculated in Section \ref{sec:corrLengths}.
		In Section \ref{sec:finite} we indicate how our results are affected if the particles have a finite aspect ratio, before concluding with the main findings of this study in Section \ref{sec:dis}.

%%%%%%%%%%%%%%%%%%%%%%%%%%%%%%%%%%%%%%%%%%%%%%%%%%%%%%

	\section{Theoretical model of the phase transition \label{sec:Onstheory}}
		We model rod-like nanoparticles as impenetrable straight spherocylinders, with a cylindrical body of length $L$ and diameter $D$, and hemispherical end-caps of the same diameter.
		The interaction potential between two particles is infinite if two particles overlap, and zero otherwise.
		A particle's orientation is characterised by the unit vector $\bu$ along its main body axis.
		It is defined relative to the direction of the external field, which corresponds to the $z$-axis of our cartesian coordinate system, so that $\bu^T= (\sin \theta \cos \varphi, \sin \theta \sin \varphi, \cos \theta )$, with $\theta$ the polar angle and $\varphi$ the azimuthal angle.
		Particle orientations are distributed according to the orientational distribution function $\psi(\bu)$, which in the isotropic phase is a constant $\psi(\bu) = (4\pi)^{-1}$.
		In the weakly ordered paranematic phase and in the strongly ordered nematic phase, the orientational distribution depends on the field strength $K$ and on the dimensionless particle concentration $c = \rho \pi L^2 D / 4 $, with $\rho$ the number density. 
		Due to the inversion symmetry and cylindrical symmetry of the paranematic and nematic phases, $\psi (\bu) = \psi (- \bu)$ and $\psi (\bu)=\psi (\theta) = \psi (\pi - \theta)$.
		
		To model field-induced particle alignment, we define a quadrupole field that aligns the particles along the $z$-axis,
		\begin{align}
			\beta U = -K \cos ^{2} \theta, \quad \text{with} \quad K>0. \label{eq:flowfield}
		\end{align}
		Here, $\beta = (k_\mathrm{B}T)^{-1}$ is the reciprocal thermal energy, with $k_\mathrm{B}$ the Boltzmann constant and $T$ the absolute temperature.
		The exact expression for the dimensionless field strength strongly depends on the particular type of field applied.
		In the case of an electric quadrupole field, it is defined as $K=\beta \Delta \alpha E^{2} / 2$, with $E$ the strength of the electric field and $\Delta \alpha$ the particle's polarisability anisotropy.
		If the field is magnetic in nature, we get $K=\beta  \Delta \chi H^{2} / 2$, with $H$ the magnetic field strength and $\Delta \chi$ the magnetic susceptibility anisotropy.
		Also a rod submerged in a thermotropic nematic host medium is subject to a quadrupole-type potential.~\cite{Burylov1990, Burylov1994}
		For carbon nanotubes in a nematic host matrix, we need to consider the limit of weak anchoring, in which case the field strength is $K=-\beta LDW\pi/3$.\footnote{Note that here we included the factor $\beta$ in the definition of $K$, which is missing in Ref.~\cite{Finner2018}.}
		Here $W$ is an anchoring energy per surface area.~\cite{PopaNita2008}
		To describe particle alignment caused by liquid flow, we focus on steady-state uniaxial elongational flow, which is the only flow field that allows for a quasi-static treatment.~\cite{Khokhlov1982, DoiEdwards}
		We assume that the network structure is frozen in fast enough upon solidification for our equilibrium description to hold.
		The field strength is then defined as $K=3 \dot{\epsilon} / 4 D_\mathrm{r}$, where $\dot{\epsilon}$ is the strain rate and $D_\mathrm{r}$ the rotational diffusion coefficient of a straight rod-like particle.
		
		To calculate the orientational distribution function, we make use of Onsager theory, which becomes exact in the limit of infinitely slender particles, $L/D\rightarrow \infty$.~\cite{Onsager1949}
		In practice, the theory becomes quantitative for aspect ratios in excess of a few hundred.~\cite{Lee1987, Frenkel1987}
		Including the contribution of our external orienting field, Eq.~\eqref{eq:flowfield}, the Helmholtz free energy per particle reads
		\begin{align}
			f(c) =\,\frac{\beta F(c)}{N}=\,& \ln c-1+ \langle \ln [\psi(\bu)] \rangle\\
				 &+\frac{4 c}{\pi} \langle \langle |\bu \times \bu'| \rangle \rangle' - K \langle \cos ^{2} \theta  \rangle, \nonumber
		\end{align}
		with $N = \rho V$ the number of particles in the volume $V$.
		Here, the angular brackets denote the orientational average $\langle \cdots \rangle = \int \dd \bu (\cdots) \psi(\bu) = \int_0^{2\pi} \dd \varphi \int_0^{\pi} \dd\theta \sin\theta (\cdots) \psi(\bu) $, with a similar definition for the primed variable.
		
		Minimising the free energy results in the self-consistent Onsager equation for the orientational distribution function, \cite{Onsager1949, Varga1998, Khokhlov1982}
		\begin{align}
			\ln \psi(\bu)= k + K \cos ^2\theta -\frac{8 c}{\pi} \int \dd \bu' \psi\left(\bu'\right)|\bu \times \bu'|, \label{Onsager}
		\end{align}
		with $k$ a Lagrange-multiplier, determined by the normalisation constraint $\int \dd \bu \psi(\bu)=1$.
		%Taking the exponent of Eq.\, \eqref{Onsager} and exploiting the symmetry of the system, the integrations can be simplified to
		%\begin{align}
		%	\psi(\theta)=\frac{1}{Z} \exp \left(K \cos ^{2} \theta-\frac{16 c}{\pi} \int_{0}^{\pi / 2} \dd \theta' \psi\left(\theta'\right) \int_{0}^{2 \pi} \dd \varphi'|\bu \times \bu'|\right). \label{Onsager2}
		%\end{align}
		After solving the Onsager equation, the alignment of particles along the $z$-axis can be quantified with the nematic order parameter
		\begin{align}
			\langle P_2 \rangle = \frac{1}{2} \langle 3\cos^2\theta -1 \rangle,
		\end{align}
		which is zero in the isotropic phase, $\langle P_2 \rangle=1$ for perfectly parallel particles and $\langle P_2 \rangle=- 1/2$ for particles perfectly perpendicular to the $z$-axis.~\cite{Odijk1986review}
		
		In the absence of a field, the isotropic orientational distribution $\psi(\bu) = (4\pi)^{-1}$ always solves the Onsager equation. 
		At low particle concentrations it is, in fact, the only solution to Eq. \eqref{Onsager}.
		For high concentrations, additional (meta-)stable solutions arise, and the isotropic solution becomes unstable.
		In this case, we have the nematic phase with an order parameter $\langle P_2 \rangle > 0$, in which particles spontaneously align along a common axis called the nematic director.
		The second metastabe solution is a disaligned phase, with particles preferentially oriented perpendicular to the director, and a negative nematic order parameter.~\cite{Odijk1986review, Vroege1992}
		
		In the presence of an orienting field with $K>0$, the isotropic orientational distribution is no longer a solution to Eq. \eqref{Onsager}.
		In this case, particles at low concentration form the weakly ordered paranematic phase and, upon density increase, exhibit a first order phase transition to the nematic, at least for weak orienting fields.
		If the field strength is sufficiently large, no phase transition occurs, as the paranematic and nematic phases are indistinguishable.~\cite{Lee1987, Varga1998, Khokhlov1982, Lettinga2004, Kityk2008}
		In the paranematic phase (P), the order parameter is typically smaller than in the nematic (N), with $0\leq\langle P_2 \rangle \leq 1$ for both phases, and we consider the isotropic phase with $K=0$ and $\langle P_2 \rangle = 0$ a special case of the paranematic.
		At the phase transition, the two phases coexist and need to be in thermal equilibrium.
		The particle concentrations $c_\mathrm{P}$ and $c_\mathrm{N}$ of an athermal system at coexistence can then be calculated by equating the dimensionless pressures $p = c^2 \dd f(c) / \dd c$ and the chemical potentials $\mu = f(c) + p(c) / c$ in the respective phases.
		
		After outlining the theory of the phase transition, we now turn to the theoretical description of cluster formation.
		
	\section{Continuum Percolation Theory \label{sec:percTheory}}
		To theoretically study clustering within a nanoparticle dispersion, it is first necessary to define a connection criterion for particle pairs.
		We can then proceed to investigate two-body distribution functions and choose only the contributions of particles that are, in some way, connected to each other.
		In geometric percolation, two particles are considered to be directly connected if their surface-to-surface distance is smaller than a cutoff distance $\lambda$, which we call the connectivity range.
		This way, we effectively model the hard body of the particle to be centered inside a spherocylindrical ``contact shell'' of diameter $D+\lambda$.
		A connection between two particles is achieved if their contact shells overlap.
		In the context of electrical percolation in polymeric nanocomposite materials, the connectivity range $\lambda$ corresponds to the effective tunneling length of charge carriers through the polymeric matrix.
		In charge-stabilised solutions, where ions are charge-carriers, it must be on the order of the Debye length.~\cite{Vigolo2005, Finner2019PRL}
		
		We now turn to the radial distribution function $g(\br, \bu, \bu')$, which measures how particles with position $\br$ and orientation $\bu$ locally order around a test particle at the origin with orientation $\bu'$.
		The radial distribution function can be separated into two contributions, one that arises from particles that are in some way connected and therefore part of the same cluster, and one that describes the correlation of particles that are \textit{not} part of the same cluster:~\cite{Torquatobook, Coniglio1977, Bug1986}
		\begin{align}
			g(\br, \bu, \bu')= P(\br, \bu, \bu') + D(\br, \bu, \bu').
		\end{align}
		Here, $P(\br, \bu, \bu')$ is the pair connectedness function, which represents the (unnormalised) probability that two particles with the orientations $\bu$ and $\bu'$ at relative distance $\br$ from each other are connected in some way. 
		The pair blocking function $D(\br, \bu, \bu')$ denotes the contribution of disconnected particles to the pair correlation.
		The cluster structure factor, representing the weight-average number of particles within a cluster on a length scale defined by the wave vector $\bq$, is then given by
		\begin{align}
			S(\bq)=1+\rho \langle \langle \hat{P}(\bq, \bu, \bu') \rangle \rangle'.
		\end{align}
		In the thermodynamic limit, the particle cluster becomes infinitely large at the percolation threshold, implying that $ \lim_{\bq\rightarrow \mathbf{0}} S(\bq) \rightarrow \infty$.
		
		The pair connectedness function $P(\br, \bu, \bu')$ can be obtained from the connectedness Ornstein-Zernike equation,~\cite{Torquatobook, Coniglio1977, Bug1986} which in Fourier space reads
		\begin{align}
			\hat{P}(\bq, \bu, \bu')=\hat{C}^+(\bq, \bu, \bu')+ \rho \langle \hat{C}^+(\bq, \bu, \bu'') \hat{P}(\bq, \bu'', \bu') \rangle''. \label{FTcOZE}
		\end{align}
		Here, the direct connectedness function $\hat{C}^+(\bq, \bu, \bu')$ is proportional to the probability that the two test particles are connected, and that their connecting path is devoid of so-called ``bottleneck particles''.
		These are particles that, upon removal from the cluster, would leave the two test particles disconnected from each other.
		The second term in Eq.\,\eqref{FTcOZE} self-consistently describes all other configurations of particle networks that do contain one or more bottleneck particles.
		Note that Eq.~\eqref{FTcOZE} explicitly accounts for angular particle correlations.
		
		To make headway, we need to insert a closure relation, \textit{i.e.}, an appropriate approximation for the direct connectedness function $C^+$.
		Within the second virial approximation closure, valid in the limit of large particle aspect ratios,  $\hat{C}^{+}\left(\bq, \bu, \bu'\right)$ is equivalent to the (Fourier transformed) connectedness Mayer function, to leading order in the particle length $L\gg D$ and for large wave lengths on the scale of the particle width, $|\bq|D \ll 1$,~\cite{Onsager1949, Balberg1984, Otten2012}
		\begin{align}
			\hat{f}^{+}\left(\bq, \bu, \bu'\right)=2 L^2\lambda\left|\bu \times \bu'\right| j_{0}\left(\frac{L \bq \cdot \bu}{2}\right) j_{0}\left(\frac{L \bq \cdot \bu'}{2}\right). \label{C3f+}
		\end{align}
		In the limit of vanishing wave vectors, the function $\hat{f}^{+}\left(\mathbf{0}, \bu, \bu'\right)$ is equivalent to the \textit{contact volume} of two particles.~\cite{Balberg1984}\\
		
		For convenience, we again introduce the dimensionless particle concentration $c=\rho \pi L^2 D / 4$, and take the limit of vanishing wave vectors to investigate the onset of percolation in the thermodynamic limit.
		Defining the (dimensionless) function
		\begin{align}
			\hat{h}\left(\bu\right)=\frac{4}{\pi D L^{2}} \langle \hat{P}\left(\mathbf{0}, \bu, \bu'\right) \rangle', \label{C3h}
		\end{align}
		we can then re-write the macroscopic average cluster size as
		\begin{align}
			 S(\mathbf{0}) &= 1+ c \langle \hat{h}(\bu) \rangle_\bu \label{dimlessS}.
		\end{align}
		The dimensionless connectedness Ornstein-Zernike equation within the second virial approximation, $\hat{C}^{+}\left(\bq, \bu, \bu'\right)=\hat{f}^{+}\left(\bq, \bu, \bu'\right)$, becomes
		\begin{align}
			\hat{h}\left(\bu\right) =\frac{8}{\pi} \frac{\lambda}{D}\langle \left|\bu \times \bu'\right|\rangle'+ \frac{8}{\pi} \frac{\lambda}{D} c \langle \left|\bu \times \bu'\right| \hat{h}\left(\bu'\right)\rangle'. \label{dimlessFTcOZE} 
		\end{align}
			
		To obtain the average cluster size \eqref{dimlessS}, Eq.~\eqref{dimlessFTcOZE} can now be solved numerically by recursive iteration, provided that the orientational distribution function is known.
		Note that, in contrast to the particle concentration $c$, the orienting field strength $K$ does not enter Eq.~\eqref{dimlessFTcOZE} directly.
		It indirectly affects the percolation threshold via the averages $\langle \cdots \rangle = \int \dd \bu (\cdots) \psi(\bu)$, as the orientational distribution function $\psi(\bu)$ is coupled to the particle concentration and the field strength through Eq.~\eqref{Onsager}.
		Before discussing our results for the percolation threshold, we describe our numerical procedure in the following Section.

%%%%%%%%%%%%%%%%%%%%%%%%%%%%%%%%%%%%%%%%%%%%%%%%%%%%%%

	\section{Numerical Methods}\label{sec:numerics}
		To study cluster formation in anisotropic phases and numerically identify regions of percolation within the phase diagram, we either perform sweeps in the particle concentration for a fixed field strength $K$, or sweeps in the field strength for a fixed particle concentration $c$.
		For each parameter combination, we first calculate the distribution function of particle orientations by recursive iteration of the Onsager equation, Eq.~\eqref{Onsager}, as described in Ref.~\cite{vanRoij2005}.
		
		Our iteration is performed on an angular grid with $N_\varphi = 400$ grid points for the azimuthal angle $0 \leq \varphi \leq 2 \pi$, and $400 \leq N_\theta \leq 1600$ grid points for the polar angle $0 \leq \theta \leq \pi/2$.
		In order to enhance the resolution around the peak of the distribution function, we (arbitrarily) divide our $\theta$-grid into three equidistant grids with $N_\theta/2$ points in the range $[0,\pi/8)$, $N_\theta/4$ points in $[\pi/8,\pi / 4)$ and  $N_\theta/4$ grid points in $[\pi/4,\pi/2)$.
		
		As a starting point for the iteration, we use either the isotropic distribution function $\psi(\theta) = (4\pi)^{-1}$, or the Gaussian distribution, $\psi(\theta)=c^2 \exp (- 2c^2 \theta^2/ \pi) / \pi^2$ for $0 \leq \theta \leq \pi / 2$, and $\psi(\pi - \theta)$ for $\pi/2 \leq \theta \leq \pi$.
		The recursive iteration of Eq.\,\eqref{Onsager} is continued until the difference between subsequent iterations of $\psi(\theta)$ at each grid point is smaller than our iteration tolerance of $10^{-7}$.
		
		Focusing on determining the percolation threshold, we then iterate the dimensionless connectedness Ornstein-Zernike equation, Eq.~\eqref{dimlessFTcOZE}, for a fixed value of the connectivity range $\lambda / D$,  to obtain a discrete representation of the function $\hat{h}(\bu)$.
		Our iteration scheme is similar to that of the Onsager equation,~\cite{vanRoij2005} and is performed on the same angular grid and with the same iteration tolerance.
		We initialise the percolation iteration with the ($c$-independent) source term of the connectedness Ornstein-Zernike equation, Eq. \eqref{dimlessFTcOZE}.
		
		If the iteration procedure diverges, and the difference between subsequent iterations grows throughout the first 5000 steps, we abort the iteration and assume a percolating network.
		If the iteration converges, particle clusters are finite, which means that no percolating network can occur.
		This method locates the boundary of the percolation region with an absolute uncertainty of half a step size, \textit{i.e.}, $\Delta K / 2 = 0.001$ in a $K$-sweep, and $\Delta c / 2 = 0.005$ in a concentration-sweep.
		To estimate the percolation boundary more accurately, we calculate the inverse of the average cluster size, $S^{-1}$ outside of the percolation region.
		The percolation threshold is then found by linearly extrapolating the last two data points outside of the percolation region to $S^{-1} =0$.
		
		To locate the binodals, \textit{i.e.}, the concentrations at phase coexistence in the paranematic (or isotropic) and in the nematic phase, we enforce thermal equilibrium and simultaneously solve the equations
		$p\left(c_{\mathrm{P}}\right)=p\left(c_{\mathrm{N}}\right)$ and $\mu\left(c_{\mathrm{P}}\right)=\mu\left(c_{\mathrm{N}}\right)	$
		using the Newton-Raphson method.~\cite{vanRoij2005, numericalrecipes}
		Every step of the Newton-Raphson procedure involves iteration of the Onsager equation and calculation of the respective pressures and chemical potentials as described in Ref.~\cite{vanRoij2005}.
		Our results for the binodals and for the percolation thresholds are presented in the following Section.

%%%%%%%%%%%%%%%%%%%%%%%%%%%%%%%%%%%%%%%%%%%%%%%%%%%%%%

	\section{Percolation across the paranematic - nematic transition \label{sec:percresults}}
		\begin{figure}
			\includegraphics[width = 0.75 \linewidth]{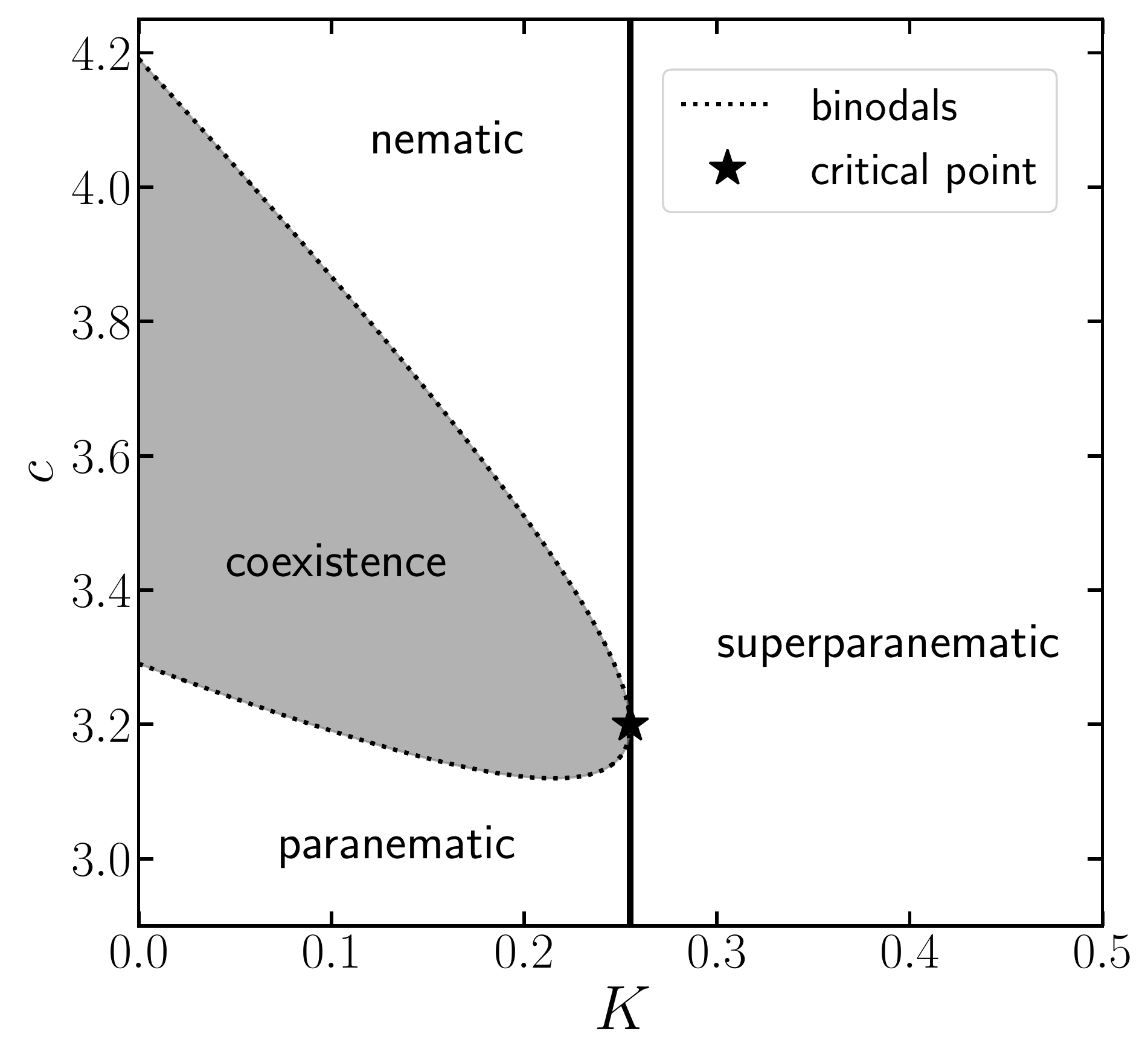}
			\caption{
				Phase diagram of the paranematic-nematic transition. Above a critical field strength $K_\text{c}\approx 0.255$, the paranematic and nematic phase are indistinguishable, with the supercritical phase denoted as \textit{superparanematic}.
				}
			\label{fig:binodals}
		\end{figure}
		In Figure \ref{fig:binodals} we show the phase diagram of the paranematic-nematic transition.
		With increasing field strength $K$, the coexistence region is found to become narrower and ends in the critical point at $(K_\text{c}, c_\text{c}) = (0.255, 3.2)$, from which onwards the paranematic and nematic phases are indistinguishable.~\cite{Lee1987, Varga1998, Khokhlov1982, Lettinga2004, Kityk2008}
		We denote the supercritical region for field strengths $K>K_\text{c}$ as \textit{superparanematic}.
		Our values of the coexistence concentrations and the critical point agree well with the literature, with relative errors of less than two percent in the critical field strength and less than $0.7\%$ in the critical concentration.~\cite{Lee1987, Varga1998}
		Using an analytical method with a variational solution of the Onsager equation instead of a numerical approach has been shown to overestimate the field strength $K_\text{c}$ at the critical point by a factor of more than 2.~\cite{Varga1998}
		
		\begin{figure}
			\includegraphics[width =\linewidth]{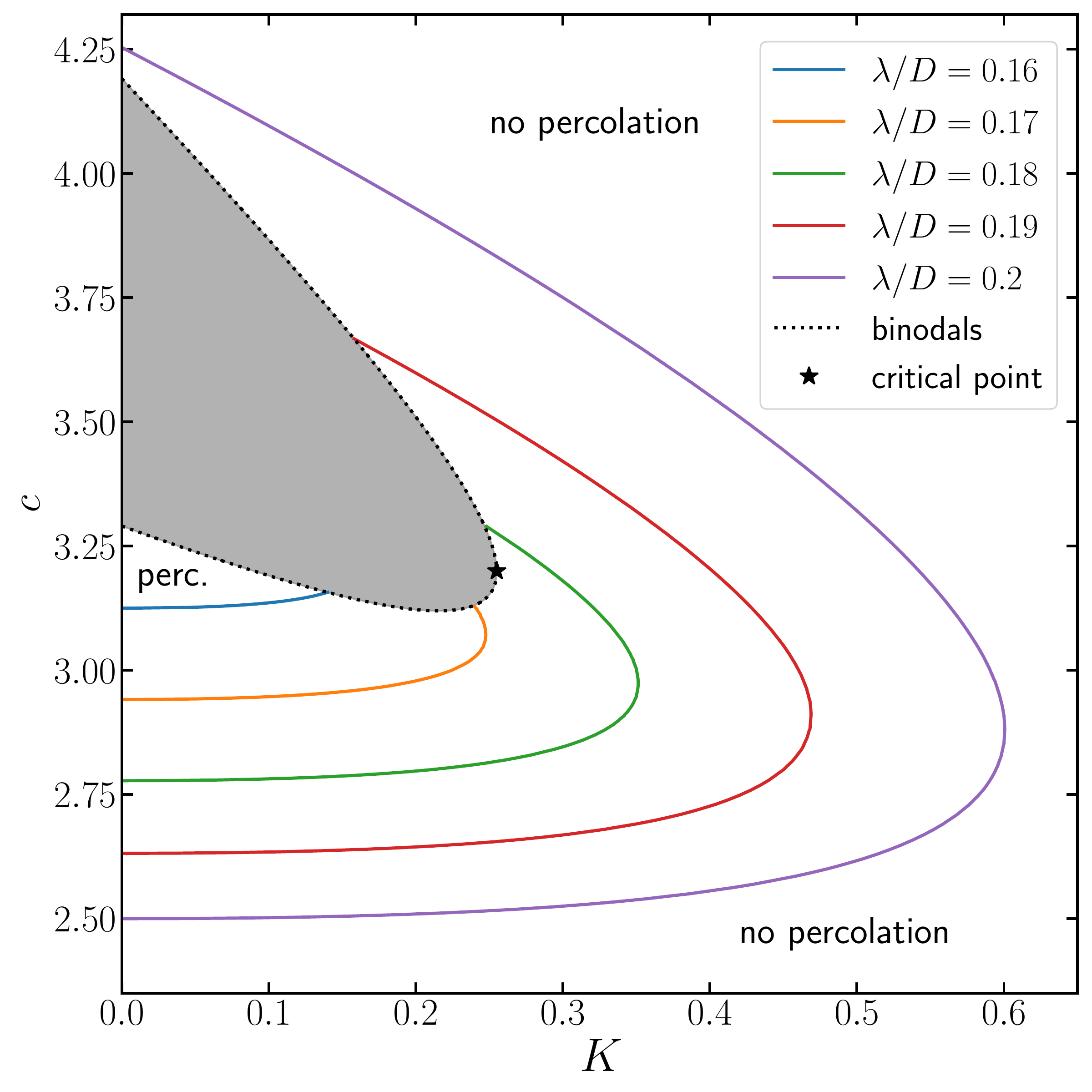}
			\caption{
				Percolation thresholds of infinitely slender particles for small values of the connectivity range $\lambda / D$.
				Within the confined percolation island \textit{i.e.}, between the percolation line and the vertical axis, a percolating network is formed.
				Outside of this region, percolation is not possible due to low particle concentrations or strong particle alignment.		
				Also indicated are the coexistence domain (grey region) and the critical point (black star).
				}
			\label{fig:droplet}
		\end{figure}
		Focusing on the percolation threshold, our numerically determined percolation thresholds for connectivity ranges between $\lambda = 0.17 D$ and  $\lambda = 0.2 D$ are shown in Figure \ref{fig:droplet}, together with the coexistence region of the phase diagram.
		The Figure demonstrates that both the particle concentration and the external field strength have a strong influence on cluster formation in the (para)nematic phase.
		We find bounded percolation ``islands'', meaning that, for weak fields, percolation is gained with increasing concentration, but can be lost again across the P-N transition or within the nematic phase if more particles are added to the suspension.
		This phenomenon of re-entrance percolation has recently been found theoretically in the absence of orienting fields, and is caused by an increase in surface-to-surface distance as the particles transition into a more (orientationally) ordered state,~\cite{Finner2019PRL} similarly to the disentanglement of rods in elongational flow fields.~\cite{Odijk1988}
		What changes for field strengths $K>0$ is that the percolation region becomes narrower and ultimately vanishes.
		This is because the additional alignment induced by the external field enhances the surface-to-surface distance of particles, thereby working against network formation.
		
		Figure \ref{fig:droplet} implies that, depending on the connectivity range $\lambda$ and the particle concentration $c$, percolation of nanoparticles in a thermotropic nematic host matrix may in practice be switched on or off by varying the \textit{temperature} of the dispersion.
		This is because the field strength $K=-\beta L D W \pi/3$ depends on the temperature via the thermal energy $\beta = 1 / \kbt$,~\cite{PopaNita2008} and also implicitly via the surface energy $W(T)$.~\cite{Faetti1985}
		For multi-walled carbon nanotubes in the thermotropic nematic matrix 5CB, for instance, the dimensionless field strength is on the order of $K=4$, and for nanotube \textit{bundles} in the polymer E7 on the order of $K=11$.~\cite{PopaNita2008, Lynch2002, Dierking2005}
		As the field strength scales roughly as $LD/ld$, where $l$ and $d$ are the length and diameter of the mesogens of the polymeric host, respectively, the range of $K$ may additionally be tuned by modifying the size ratio of the host- \textit{vs.} filler-particles.~\cite{PopaNita2008}\\
		
\begin{figure}
			\includegraphics[width = \linewidth]{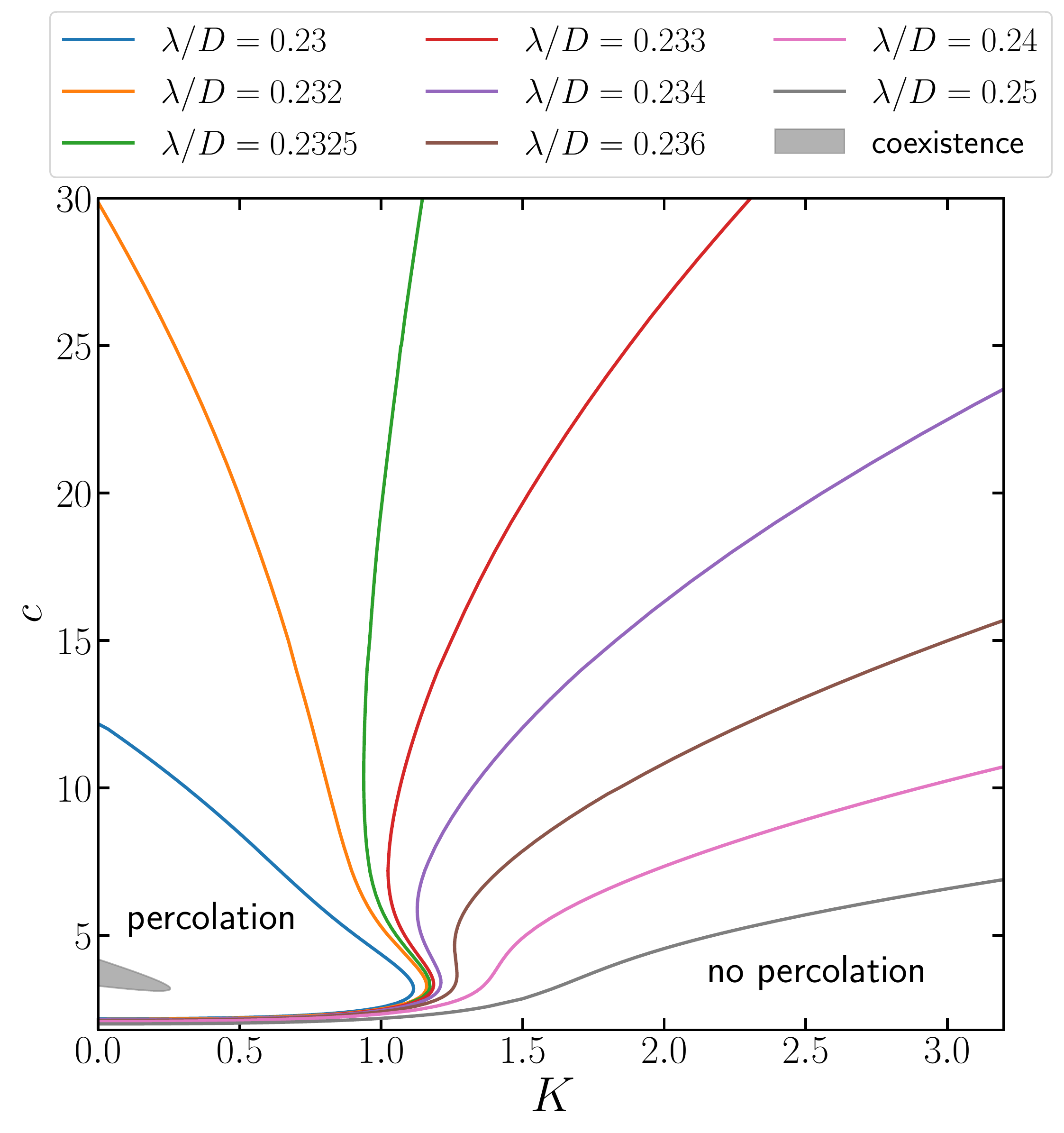}
			\caption{
				Percolation thresholds of infinitely slender particles for several connectivity ranges $\lambda / D$.
				In the region between the percolation line and the vertical axis, a percolating network is formed, while outside of that region percolation does not occur.
				With increasing connectivity range $\lambda / D$, the confined percolation island becomes larger, deforms, and eventually opens up to become an unbounded percolation area.
				The deformation of the percolation island causes multiple re-entrance effects around a field strength of $K \approx 1.15$, where percolation can subsequently be gained, lost, gained, and ultimately lost again (not shown in the plot) upon increasing the particle concentration.
				For connectivity shells larger than $\lambda / D \approx 0.236$, only one percolation threshold exists at a fixed field strength $K$, and percolation can no longer be lost with increasing particle concentration.~\cite{Finner2019PRL}
			}
			\label{fig:opening}
		\end{figure}

		Our percolation islands found in Figure \ref{fig:droplet} are similar to the results of an earlier theoretical study by Otten \etal,~\cite{Otten2012} who used an analytical expansion of the contact volume in Legendre-polynomials to predict the percolation threshold.
		However, in Ref.~\cite{Otten2012}, re-entrance takes place in the weakly ordered paranematic phase, while in our study it typically only occurs in the nematic phase, \textit{i.e.}, at much higher particle concentrations.
		An exception occurs for connectivity ranges around $\lambda / D = 0.17$, for which we do find re-entrance in the paranematic phase in a very small parameter range, as Figure \ref{fig:droplet} shows.
		While the re-entrance concentration for $K=0$ in Ref.~\cite{Otten2012} does not seem to depend on the connectivity range $\lambda / D$, in our numerical study it does.
		Also, Otten \etal~find the island shape to be present for a wide range of contact shell thicknesses of at least $0.3 \leq \lambda / D \leq 1$.
		Increasing the connectivity range to values larger than 0.232, however, we find a qualitatively different percolation diagram, which we present in Figure \ref{fig:opening}.
		
		Figure \ref{fig:opening} shows that, with an increase of the connectivity range, the percolation behaviour changes drastically, and the bounded percolation island opens up to an infinitely large percolation domain.
		This means that, for large enough connectivity ranges, percolation can no longer be lost with increasing concentration, only gained.
		In this case, the addition of particles is able to compensate for the additional translational freedom that the enhanced ordering provides.
		We surmise that the qualitative difference between our numerical results for the percolation regions and the analytical prediction in Ref.~\cite{Otten2012} arise from their analytical expansion in Legendre polynomials to third order, which seems to be insufficient for very strong particle alignment.
		
		It turns out that, close to the opening transition, the percolation island deforms in such a way that re-entrance can be observed multiple times. 
		For $\lambda = 0.233D$ at a field strength of $K=1.15$, for instance, percolation can be first gained, then lost, and then gained again with increasing particle concentration.
		Note that our analytical results in the absence of a field suggests that the percolating network must ultimately be destroyed again at higher concentrations (not shown in the plot).
		In fact, all curves for connectivity ranges below a critical $\lambda / D \approx 0.2368$ are expected to return to the vertical axis and thereby close the percolation island at very high particle concentrations.~\cite{Finner2019PRL}

\begin{figure}
			\includegraphics[width = \linewidth]{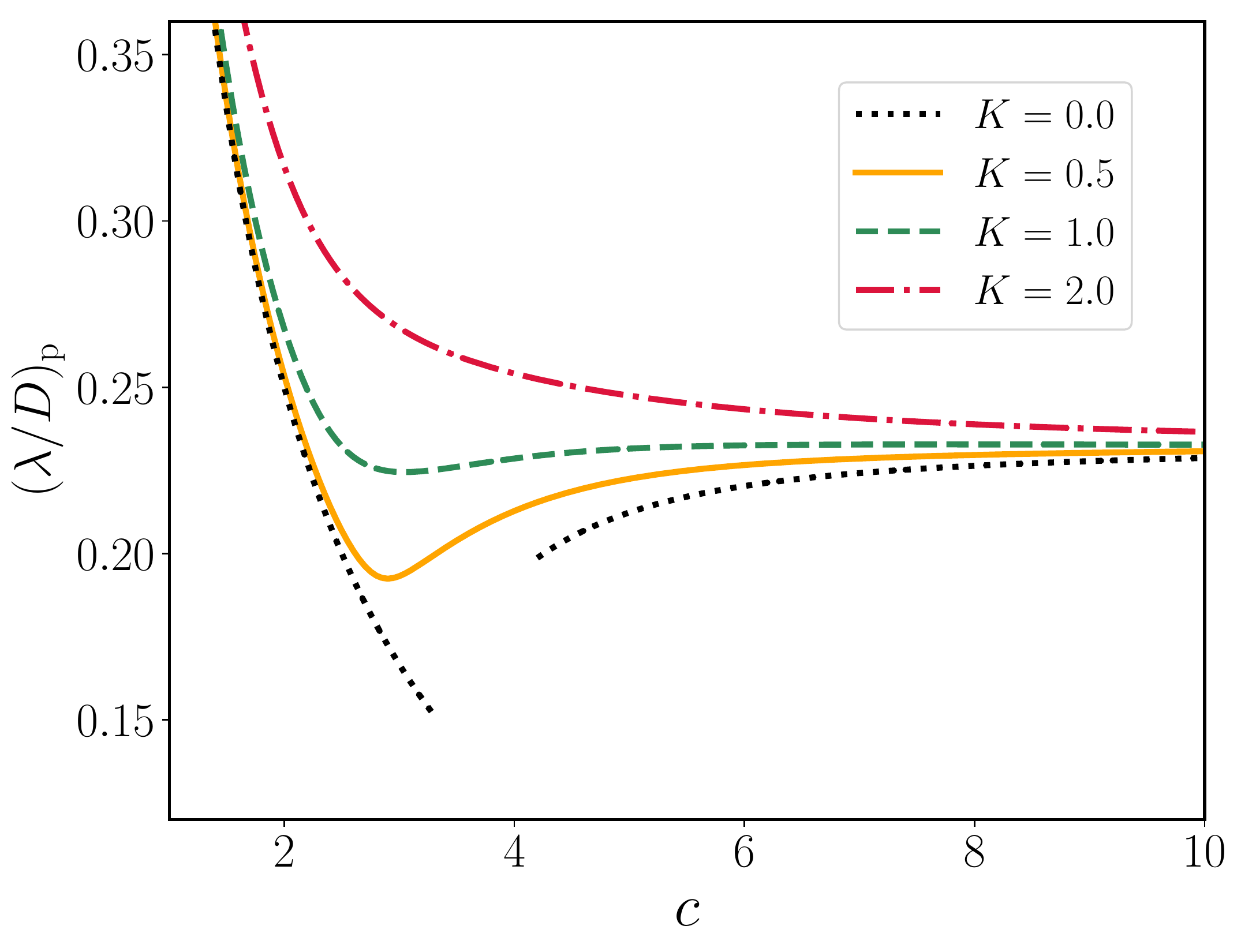}
			\caption{
				Critical connectivity range $(\lambda / D)_\pp$ required for percolation as a function of the particle concentration $c$ for various orienting field strengths $K$.
				At low concentrations, the field has an adverse effect on cluster formation, and determines whether percolation and re-entrance can be observed.
				For large particle concentrations, the effect of the external field becomes negligible.
				Note that for $K=0$ the transition is first order, whilst for the other values of the field strength the dispersion is superparanematic.
			}
			\label{fig:lambdas}
		\end{figure}

		To further investigate how the orienting field affects percolation, we fix the field strength $K$ and calculate the concentration dependence of the critical connectedness shell thickness needed for percolation.
		Our results are shown in Figure \ref{fig:lambdas} for different values of $K$, and compared to the known zero-field behaviour.~\cite{Finner2019PRL}
		We find that, at low concentrations, the field works against network formation and strongly affects the percolation threshold in terms of $\lambda / D$, while its influence becomes much weaker at high particle concentrations.
		This is because the particles in the nematic are so strongly aligned by excluded volume interactions that the additional alignment induced by the external field becomes negligible.
		The possibility of finding re-entrance percolation is indicated in Figure \ref{fig:lambdas} by the nonmonotonicity of the curves for low values of the field strength.
		As the field strength is increased, the curve becomes monotonic and re-entrance breaks down.
		
		In this Section, we have shown that the alignment of particles by external or molecular fields has an immense influence on the percolation threshold.
		In view of potential applications in nanocomposite materials, however, also the structure and physical dimensions of particle networks turn out to be of importance, as they can be tuned to create materials with anisotropic transport properties.
		For the purpose of investigating the physical extent of particle clusters, we calculate the correlation lengths parallel and perpendicular to the alignment axis in the following Section.

%%%%%%%%%%%%%%%%%%%%%%%%%%%%%%%%%%%%%%%%%%%%%%%%%%%%%%pair connectedness function

	\section{Correlation lengths \label{sec:corrLengths}}
	In order to investigate the physical dimensions of particle clusters and detect possible anisotropies in the cluster shape, we need to probe the wave-vector dependence of the cluster structure factor
		\begin{align}
			S(\bq) = 1 +  \rho \langle \langle  \hat{P}(\bq, \bu, \bu'\rangle \rangle'.
		\end{align}
		For this purpose, we separate the (orientationally averaged) pair connectedness function into an isotropic and a (possibly) anisotropic contribution, $\langle \langle  \hat{P}(\bq, \bu, \bu'\rangle \rangle'= \langle \langle  \hat{P}(\bnod, \bu, \bu'\rangle \rangle'+  \hat{P}_{\mathrm{an}}(\bq)$.
		The correlation lengths, which are a measure for the physical dimensions of sub-percolating clusters, can then be found by calculating $ \hat{P}_{\mathrm{an}}(\bq)$ for small $\bq$, and determining how the pair connectedness function decays with increasing wave vector.
		This can be done by expanding $\langle \langle  \hat{P}(\bq, \bu, \bu'\rangle \rangle'$ to second order around $\bq = \bnod$, differentiating the connectedness Ornstein-Zernike equation twice and applying the second virial approximation.

		We start by expanding the pair connectedness function around $\bq=\bnod$ to second order in the wave vector,
			{\small
			\begin{align}
				\hat{P}\left(\bq, \bu, \bu'\right)=&\hat{P}\left(\bnod, \bu, \bu'\right)
						+\left.\frac{\partial \hat{P}(\bq, \bu, \bu')}{\partial \bq}\right|_{\bq=\bnod} \cdot \bq\\
						&+\frac{1}{2}\left.\frac{\partial^{2} \hat{P}(\bq, \bu, \bu')}{\partial \bq \partial \bq}\right|_{\bq=\bnod} : \mathbf{q q}+\ldots. \nonumber
			\end{align}
			}%
			Here, the linear term vanishes due to the inversion symmetry of the problem.
			The leading order term of the anisotropic part is then given by
			\begin{align}
				 \hat{P}_{\mathrm{an}}(\bq) = \frac{1}{2} \left\langle \bM(\bu) \right\rangle : \bq\bq, \label{San}
			\end{align}
			with the ($3\times 3$)-matrix
			\begin{align}
				\bM(\bu)  = \left\langle \frac{\partial^{2}}{\partial \bq \partial \bq} \hat{P}\left(\bq, \bu, \bu'\right)\Big|_{\bq=\bnod}\right\rangle'  \label{M}
			\end{align}
			
			The goal is now to find an equation for the matrix $\bM(\bu)$.			
			Differentiating the connectedness Ornstein-Zernike equation twice and evaluating it at $\bq =\bnod$ produces the equation
			{\small
			\begin{align}
				\frac{\partial^{2}}{\partial \bq \partial \bq} \hat{P}&\big(\bq,\bu, \bu'\big)\Big|_{\bq=\bnod}= \frac{\partial^{2}}{\partial \bq \partial \bq} \hat{C}^+\left.\left(\bq, \bu, \bu'\right)\right|_{\bq=\bnod}\\
				& + \rho \int \dd \bu'' \psi\left(\bu''\right) \frac{\partial^{2}}{\partial \bq \partial \bq} \hat{C}^+\left.\left(\bq, \bu, \bu''\right)\right|_{\bq=\bnod} \hat{P}\left(\bnod, \bu'', \bu'\right)\nonumber \\
				& +  \rho \int \dd \bu'' \psi\left(\bu''\right) \hat{C}^+\left(\bnod, \bu, \bu''\right) \frac{\partial^{2}}{\partial \bq \partial \bq} \hat{P}\left.\left(\bq, \bu'', \bu'\right)\right|_{\bq=\bnod}, \nonumber
			\end{align}
			}%
			where the first derivative of $\hat{P}$ again drops out for symmetry reasons.
			
			We now expand the direct connectedness function to second order in the wave vector
			{\small
			\begin{align}
				\hat{C}\left(\bq, \bu, \bu'\right)=\hat{C}\left(\bnod, \bu, \bu'\right)
						+\frac{1}{2}\left.\frac{\partial^{2} \hat{C}(\bq, \bu, \bu')}{\partial \bq \partial \bq}\right|_{\bq=\bnod} : \mathbf{q q}+\ldots ~, \nonumber
			\end{align}
			}%
			where the first derivative vanishes due to symmetry.
			Subsequently, we insert the second virial approximation $\hat{C}^{+} = \hat{f}^{+}$ and make use of the Taylor expansion $j_0(x) = 1 - x^2/6 + \cdots$ in the expression for the connectedness Mayer function, Eq.~\eqref{C3f+}.
			Using the limits
			\begin{gather}
				\lim _{x \rightarrow 0} j_{0}(x)=1, \quad \lim _{x \rightarrow 0} j_{0}'(x)=0, \quad \lim _{x \rightarrow 0} j_{0}''(x)=-\frac{1}{3},
			\end{gather}
			we find
			\begin{align}
				\frac{\partial^{2}}{\partial \bq \partial \bq} \hat{C}^+\left.\left(\bq, \bu, \bu'\right)\right|_{\bq=\bnod}=-\frac{\lambda L^{2}}{6} \left|\bu \times \bu' \right|\left(\bu \bu+\bu' \bu'\right). \label{C3C+2}
			\end{align}
			This results in the matrix equation
			\begin{align}
				\bM(\bu) = \br(\bu) + \mathbf{s}(\bu) + \mathbf{t}[\bM(\bu)] \label{McOZE},
			\end{align}
			with
			{\small
			\begin{gather}
				 \mathbf{r}(\bu) =-\frac{1}{6} \lambda L^{4} \int \dd \bu' \psi\left(\bu'\right)\left|\bu \times \bu' \right|\left(\bu \bu+\bu' \bu'\right), \\
				 \mathbf{s}(\bu) =-\frac{1}{6} \lambda L^{4}  \rho \int \dd \bu' \psi\left(\bu'\right)\left|\bu \times \bu' \right|\left(\bu \bu+\bu' \bu'\right) \langle \hat{P}(\bnod, \bu, \bu') \rangle, \\
				\mathbf {t}[\bM(\bu)] =2 \lambda L^{2}  \rho \int \dd \bu' \psi\left(\bu'\right)\left|\bu \times \bu' \right| \bM(\bu').
			\end{gather}
			}%
			Making use of the rotational symmetry around the $z$-axis and preaveraging the $\varphi$-integration of the dyadics $\bu\bu$ and $\bu'\bu'$, leaves us with only two distinct nonzero matrix elements, $M_{11}=M_{22}$ and $M_{33}$, which define the anisotropic part of the pair connectedness function as
			\begin{align}
				\hat{P}_{\mathrm{an}}\left(q_{\perp}, q_{ \|}\right) &=\frac{1}{2} \left\langle M_{11}(\bu)\right\rangle q_\perp^2 +\frac{1}{2} \left\langle M_{33}(\bu)\right\rangle q_\parallel^2.
			\end{align}
			As a result, the decay of connectedness correlations with increasing (but small) wave vector takes the functional form~\cite{Otten2012}
			\begin{align}
				\frac{\langle \langle  \hat{P}(\bq, \bu, \bu'\rangle \rangle'}{\langle \langle  \hat{P}(\bnod, \bu, \bu'\rangle \rangle'} = 1 - \xi^2_\parallel q^2_\parallel - \xi^2_\perp q^2_\perp,
			\end{align}
			which defines the correlation lengths, $\xi_{\|}$ and $\xi_{\bot}$, parallel and perpendicular to the director.
			Here, $q_\parallel = q_z$ is the wave number along the director, and $q_\perp = \sqrt{q^2_x + q^2_y}$ that in the perpendicular direction.
			The correlation lengths are then given by
			\begin{align}
				\left( \frac{\xi_{\perp}}{L}\right)^2 &= - \frac{1}{2} \frac{\left\langle M_{11}(\bu)\right\rangle}{\langle \langle \hat{P}(\bnod, \bu, \bu' )\rangle \rangle'}\\
				\shortintertext{and}
				\left( \frac{\xi_{\parallel}}{L}\right)^2 &= - \frac{1}{2} \frac{\left\langle M_{33}(\bu)\right\rangle}{\langle \langle \hat{P}(\bnod, \bu, \bu' )\rangle \rangle'}.
			\end{align}
			
			For numerical purposes, it is instructive to make the relevant equations dimensionless.
			We therefore again resort to the function
			\begin{align}
				\hat{h}\left(\bu\right)=\frac{4}{\pi D L^{2}} \langle \hat{P}\left(\bnod, \bu, \bu'\right) \rangle',
			\end{align}
			as already introduced in Eq.~\eqref{C3h}.
			Defining the renormalised matrix $\widetilde{\bM}(\bu) = 6\bM / \lambda L^4$, we obtain the dimensionless equation
			\begin{align}
				\widetilde{\bM}(\bu) = \widetilde{\br}(\bu) + \widetilde{\mathbf{s}}(\bu) + \widetilde{\mathbf{t}}[\widetilde{\bM}(\bu)] \label{dimlessMcOZE},
			\end{align}
			with		
			\begin{align}
				 \widetilde{\br}(\bu) &=- \langle \left|\bu \times \bu' \right|\left(\bu \bu+\bu' \bu'\right) \rangle'\\
				 \widetilde{\mathbf{s}}(\bu) &=- c \langle\left|\bu \times \bu' \right|\left(\bu \bu+\bu' \bu'\right) \hat{h}\left(\bu'\right)\rangle' \\
				 \widetilde{\mathbf{t}}[\widetilde{\bM}(\bu)] &=\frac{8 c}{\pi} \frac{\lambda}{D} \langle\left|\bu \times \bu' \right| \widetilde{\bM}(\bu)\rangle'.
			\end{align}
			The correlation lengths
			\begin{align}
				\left( \frac{\xi_{\perp}}{L}\right)^2\!\! = - \frac{1}{3 \pi} \frac{\lambda}{D}  \frac{\left\langle \widetilde{M}_{11}(\bu)\right\rangle}{\langle \hat{h}(\bu) \rangle}\\
				\shortintertext{and}
				\left( \frac{\xi_{\parallel}}{L}\right)^2\!\! = - \frac{1}{3 \pi} \frac{\lambda}{D}  \frac{\left\langle \widetilde{M}_{33}(\bu) \right\rangle}{\langle \hat{h}(\bu) \rangle}
			\end{align}
			can now straightforwardly be calculated by recursive iteration of $\hat{h}(\bu)$ in Eq.~\eqref{dimlessFTcOZE}, and of the two relevant matrix elements, $\widetilde{M}_{11}$ and $\widetilde{M}_{33}$, defined by Eq.~\eqref{dimlessMcOZE}.
			The numerical iteration procedure we employ is similar to the one used to solve the Onsager equation.~\cite{vanRoij2005}

		\begin{figure}
			\includegraphics[width = 0.85 \linewidth]{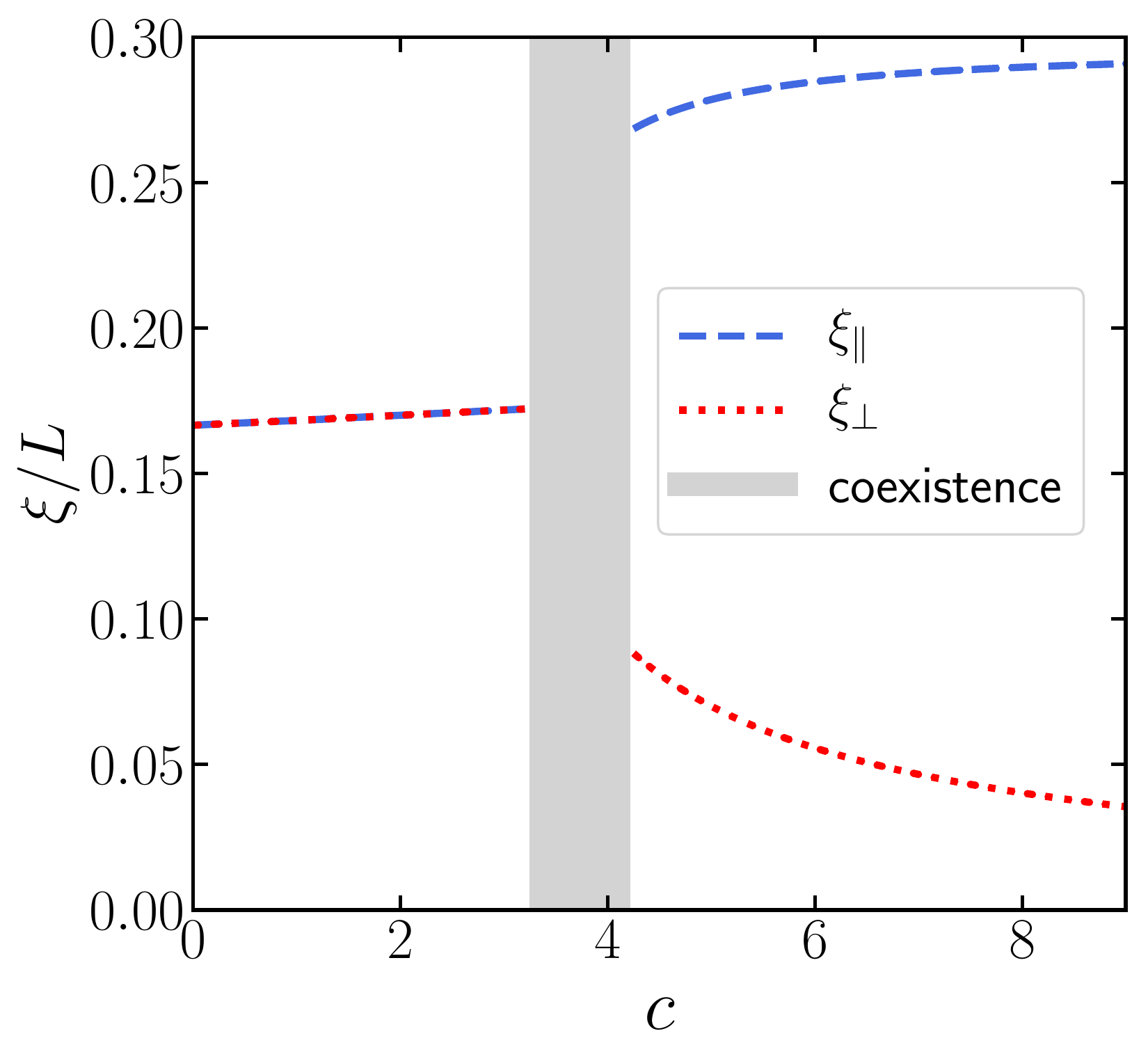}
			\caption{
				Parallel and perpendicular correlation lengths in the isotropic phase in zero field, for the connectivity range $\lambda / D = 0.01$.
				In the isotropic phase, the percolation lengths are equal, while in the nematic the correlation length along the nematic director is always larger than that in the perpendicular direction.
				Also indicated are the isotropic and nematic coexistence concentrations (binodals).
			}
			\label{fig:corrLengths1}
		\end{figure}
		
		Figure \ref{fig:corrLengths1} shows our results for the correlation lengths for the connectivity range $\lambda / D = 0.01$ in the absence of an external field.
		In this case, the correlation lengths remain finite, as the connectivity range is too small to induce percolation. 
		While both correlation lengths are equal in the isotropic phase, we find the isotropic-nematic transition to abruptly elongate the clusters.
		In the nematic, the correlation length along the nematic director is always larger than that in the perpendicular direction.
		In agreement with our previous results,~\cite{Finner2019PRL} $\xi_\parallel$ saturates to a constant value, while $\xi_\perp$ decays to zero with increasing particle concentration.
		In the limit $c  \rightarrow 0$, both correlation lengths decay to a value of $\xi_\parallel = \xi_\perp = L/6$, contradicting the earlier results of Otten \etal, who determined this value to be $L/\sqrt{12}$, \textit{i.e.}, the radius of gyration of a rod.~\cite{Otten2012}
		Note that we would expect a value of $L/\sqrt{12}$ if the theory were centred around segment densities, not rod densities, as segments within a single rod are part of the same cluster (the rod).

		\begin{figure}
			\includegraphics[width = 0.85 \linewidth]{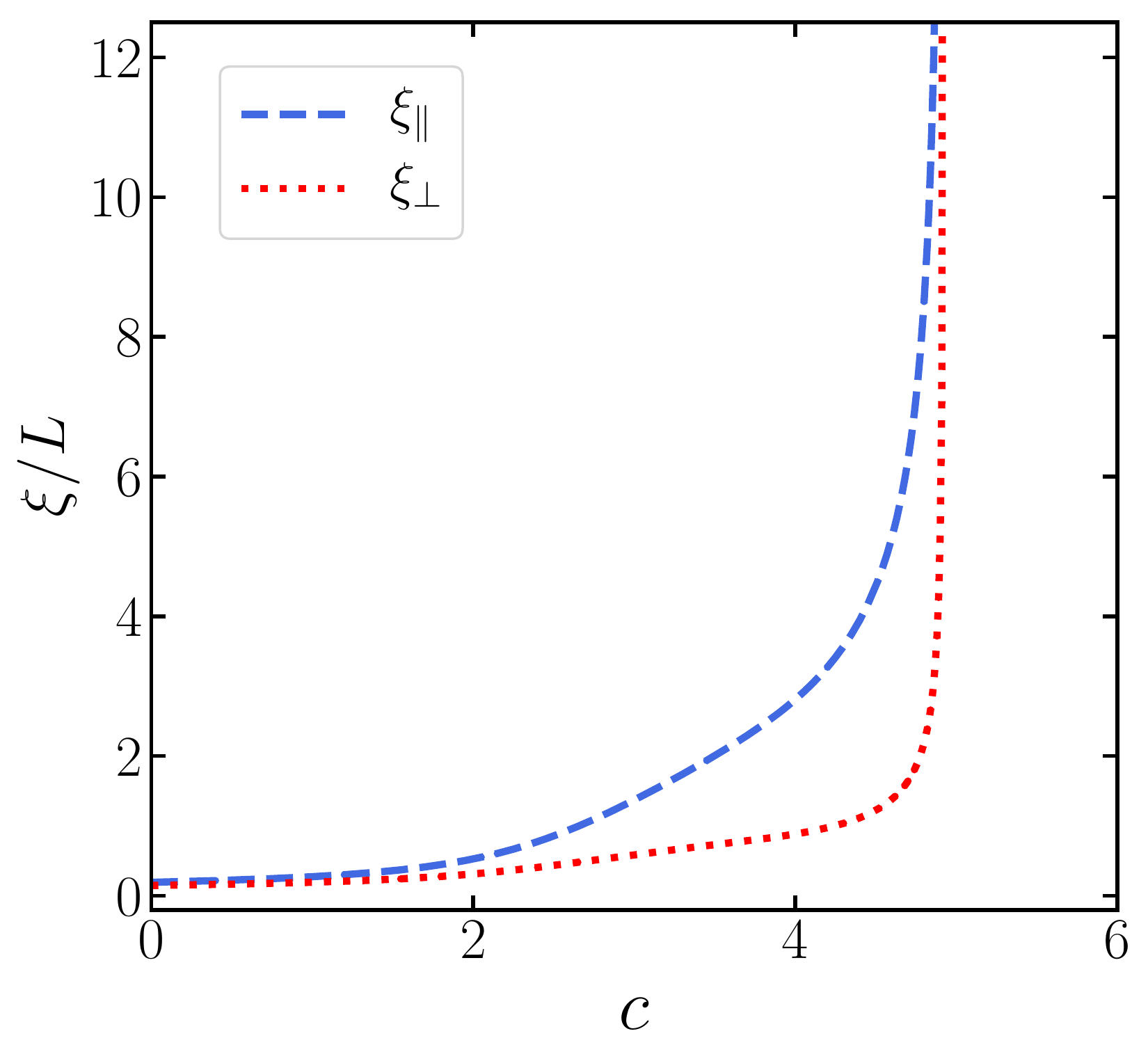}
			\caption{
				Parallel and perpendicular correlation lengths, with a connectivity range $\lambda / D = 0.24$ and a field strength $K = 1.5$ in the superparanematic regime.
				The correlation length along the field direction is larger than that in the radial direction, confirming that sub-percolating clusters in the paranematic or superparanematic phase are (weakly) elongated.
				Both correlation lengths diverge with the same critical exponent at the same concentration, which corresponds to the percolation threshold, as predicted by Otten \etal.~\cite{Otten2012}
			}
			\label{fig:corrLengths2}
		\end{figure}
\begin{figure}
			\includegraphics[width = 0.85 \linewidth]{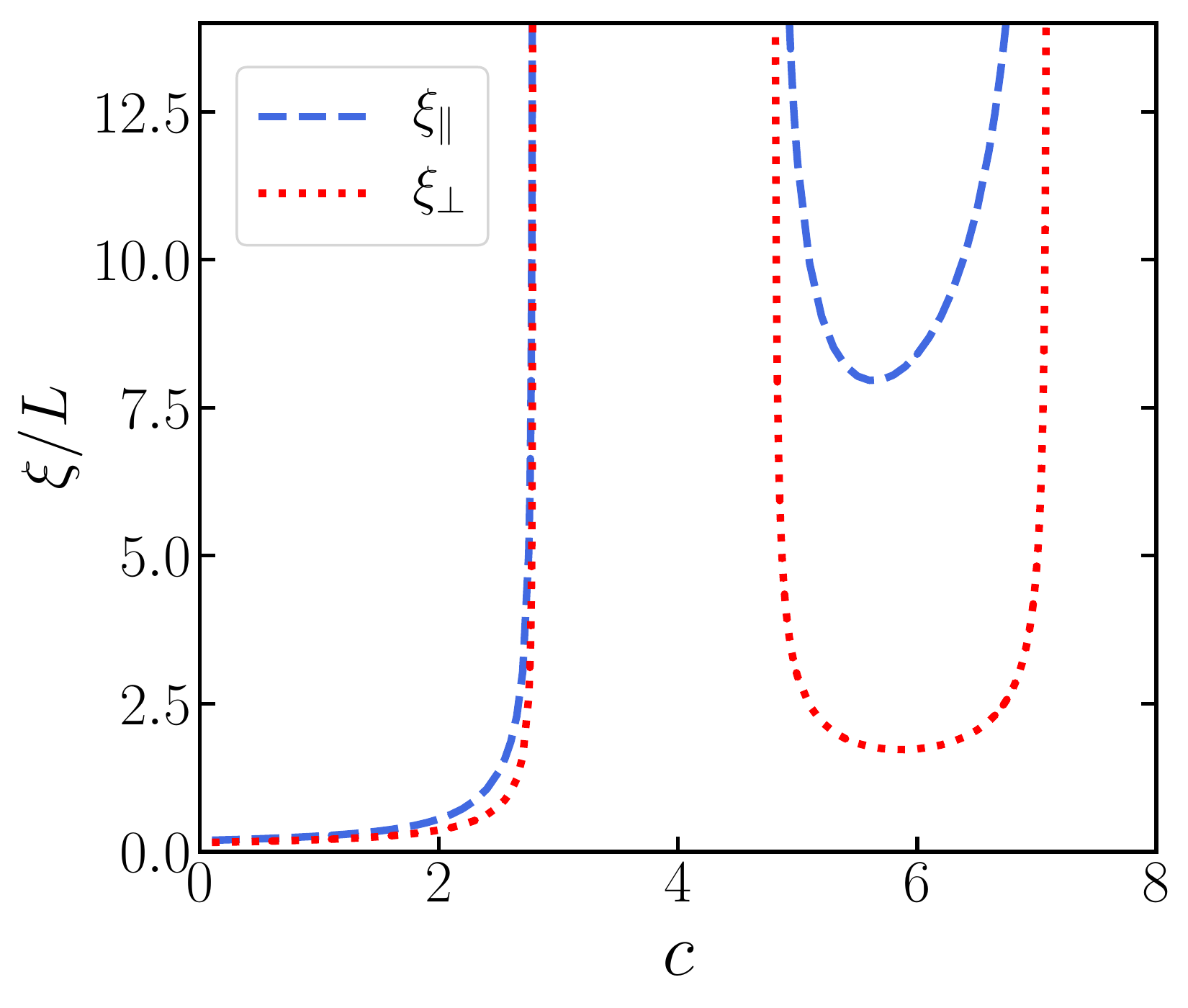}
			\caption{
				Correlation lengths parallel to the field direction (blue, dashed) and perpendicular to the field (red, dotted) as a function of the particle concentration $c$ for the field strength $K = 1.15$ and connectivity range $\lambda / D = 0.234$.
				The Figure demonstrates the re-entrance behaviour of the superparanematic dispersion: a percolating network is formed for concentrations $2.8 \leq c \leq 4.8$, but breaks down with increasing particle concentration. 
				It is formed again for $c> 7.1$.
				Both correlation lengths diverge at the percolation thresholds with different prefactors, demonstrating that particle clusters in the superparanematic phase can be weakly or strongly anisotropic.
			}
			\label{fig:corrLengths3}
		\end{figure}

		In Figure \ref{fig:corrLengths2}, we plot the correlation lengths of a superparanematic dispersion, with $\lambda / D = 0.24$ and $K = 1.5$.
		It turns out that the external orienting field causes minor elongation of the clusters, with the correlation length parallel to the field again larger than that in the radial direction.
		Despite the apparent cluster elongation, both correlation lengths diverge at the same percolation threshold~\cite{Otten2012, Kale2016} with the same critical exponent, in agreement with the approximate analytical results of Ref.~\cite{Otten2012}.
		The phenomenon of re-entrance percolation is shown in Figure \ref{fig:corrLengths3} for the example of a field strength $K = 1.15$ and connectivity range $\lambda / D = 0.234$.
		We identify two concentration regions of infinite correlation lengths, and two regions in which the cluster dimensions remain finite.
		In the paranematic phase, the orienting field only causes a slight elongation of the particle clusters, whereas in the (field-enhanced) nematic phase clusters are strongly anisotropic.
		
%%%%%%%%%%%%%%%%%%%%%%%%%%%%%%%%%%%%%%%%%%%%%%%%%%%%%%

	\section{Finite aspect ratios \label{sec:finite}}
		The question arises whether our predictions for the percolation re-entrance effects in the limit of infinite aspect ratios persist for particles with a finite aspect ratio.
		In order to adapt our percolation model to particles with aspect ratios that do not exceed a few hundred, we need to include the effects of the hemispherical particle end-caps and take into account higher than two-body interactions, which become increasingly important the shorter the particles are.~\cite{Straley1973, Mulder1985, Mulder1987, Frenkel1987}
		 Rather than including many-body interactions explicitly, we make use of a renormalised second virial approximation based on the framework of Scaled Particle Theory.~\cite{Finner2019PRE, TuinierBook, Cotter1977}
		
		According to Scaled Particle Theory, the distribution function of particle orientations in an external quadrupole field is determined by the self-consistent Onsager-like equation
		\begin{align}
			\ln \psi (\bu) =& k  + K \cos ^2\theta  \label{OnsagerSPT}\\
			&-  \frac{8}{\pi} \Gamma_\text{SPT}(\phi, L/D) c  \int \dd \bu' \psi(\bu') |\bu \times \bu'|. \nonumber
		\end{align}
		Here, the factor
		\begin{align}
			\Gamma_\text{SPT}( \phi, L / D ) = \frac{1}{1-\phi} \Big[ 1 + \frac{2+ 2 L / D}{2 + 3 L / D} \frac{\phi}{1-\phi} \Big] \label{GammaSPT}
		\end{align}
		is a function of the volume fraction of particles in the suspension, $\phi = (D/L+2D^2/3L^2)c = \rho \pi D^2 [3L + 2D]/12$, and of the particle aspect ratio $L / D$.~\cite{TuinierBook}
		
		Within our recently proposed scaled-particle percolation closure,~\cite{Finner2019PRE} and in the limit of vanishing wave vector, the direct connectedness function can be approximated by
		\begin{align}
			\hat{C}^+(\mathbf{0}, \bu, \bu') = \Gamma_\text{SPT}(\phi, L/D) \hat{f}^+(\mathbf{0}, \bu, \bu'). \label{SPTclosure}
		\end{align}
		The renormalisation factor $\Gamma_\text{SPT}( \phi, L / D )$ effectively rescales the full contact volume of a particle pair,
		\begin{align}
			\hat{f}^+(\mathbf{0}, \bu,\bu') =\, & 2L^2 \lambda |\bu \times \bu'| + 2\pi L \left[ (D+\lambda)^2 - D^2\right] \nonumber \\
			&+\frac{4\pi}{3}\left[ (D+\lambda)^3 - D^3\right], \label{f0}
		\end{align}
		which now includes the contributions of the hemispherical particle end-caps. ~\cite{Coniglio1977, Onsager1949, Balberg1984, Balberg1984_2}
		
		\begin{figure}
			\includegraphics[width = \linewidth]{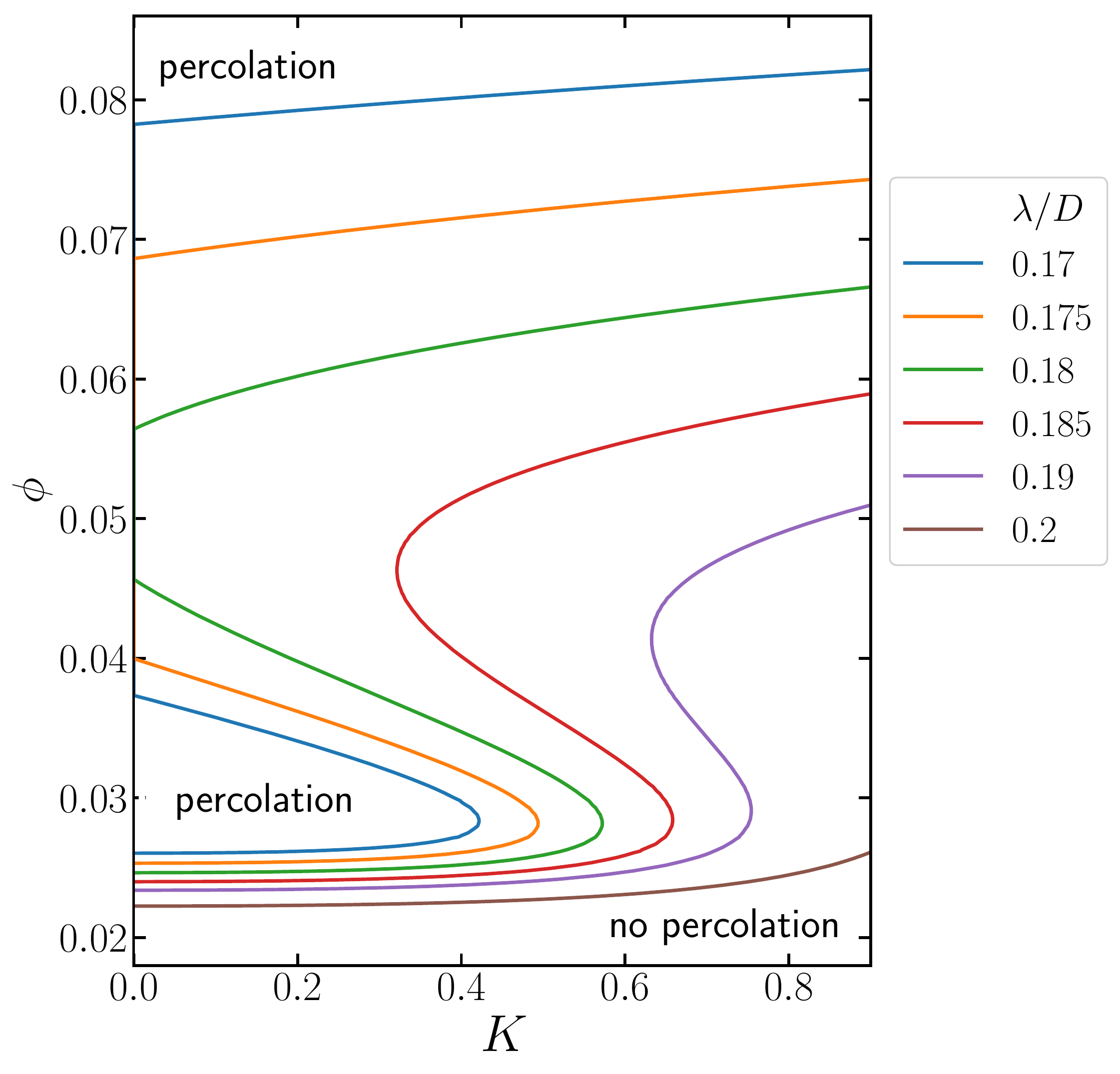}
			\caption{
				Percolation thresholds of spherocylinders with aspect ratio $L/D = 100$ for different connectivity ranges $\lambda / D$, obtained with Scaled Particle Theory in combination with the Scaled Particle percolation closure, Eq.\,\eqref{SPTclosure}.~\cite{Finner2019PRE}
				In the region between the percolation line and the $\phi$-axis, a percolating network is formed.
				Outside of that region, percolation does not occur.
				Similarly to Fig. \ref{fig:opening}, there is a bounded ``percolation island'' for small values of $\lambda / D$, which opens up with an increase of the connectivity range.
				For particles of \textit{finite} aspect ratio there exists also an additional high-density percolation threshold next to the bounded island, above which percolation can no longer be lost.
				The ``S-shape'' of the percolation curve is more pronounced than in the infinitely slender rod limit, increasing the range of field strengths for which double re-entrance can be observed.
			}
			\label{fig:finite}
		\end{figure}

		Our results for an example of particles of aspect ratio $L/D=100$ are summarised in Figure \ref{fig:finite}.
		We find that the double re-entrance effect, which is rather subtle for infinite aspect ratios, now takes place for a much larger range of field strengths $K$.
		If the connectivity shell is sufficiently thin, the ``S-shape'' of the percolation curve becomes very wide and may touch the vertical axis, producing the contained percolation island already known from the case of infinitely slender particles.
		What is different for finite aspect ratios is that there is also an additional high-density percolation threshold next to the bounded island, above which percolation can no longer be lost.
		For very low connectivity ranges, we find that the percolation island disappears entirely, leaving us with a high-density percolation threshold only (not shown in the Figure).
		These predictions are consistent with our earlier results in the absence of an external field.~\cite{Finner2019PRE}
		Percolation diagrams for the aspect ratios 10, 20 and 50 exhibit the same characteristics as Figure \ref{fig:finite} and are not shown for reasons of brevity.

%%%%%%%%%%%%%%%%%%%%%%%%%%%%%%%%%%%%%%%%%%%%%%%%%%%%%%

	\section{Discussion and conclusions \label{sec:dis}}
		In this article, we have numerically investigated the interplay between intrinsic and externally driven particle alignment on the geometric percolation of hard slender particles.
		For this purpose, we combined Onsager-theory for the paranematic-nematic transition with Continuum Percolation Theory describing the formation of particle clusters.
		For small connectivity ranges, we find that percolation may be gained with increasing concentration, but can be lost again upon particle addition, at least for weak orienting fields.
		If the orienting field is strong enough, it prevents the formation of a percolating network altogether.
		This results in a closed percolation region in the $c(K)$-phase diagram, outside of which percolation is not possible.
		
		For larger connectivity ranges, the closed ``percolation island'' strongly deforms and ultimately opens up to an unbounded domain, consistent with our earlier predictions in the absence of an external field.~\cite{Finner2019PRL}
		The deformation of the percolation island prior to its opening transition gives rise to peculiar double re-entrance effects, meaning that a percolating network may form and decay twice upon an increase of the particle concentration.
		
		Comparing our percolation diagrams with the results of an earlier analytical study by Otten \etal,~\cite{Otten2012} we observe some quantitative and qualitative differences.
		While Ref.~\cite{Otten2012} predicts the closed percolation island for a range of contact shell thicknesses of at least $0.3 \leq \lambda / D \leq 1$, our more comprehensive numerical studies here, and in an earlier Letter~\cite{Finner2019PRL}, demonstrate that the percolation region of the phase diagram must open up for connectivity ranges larger than $\lambda / D \approx 0.2368$.
		Also, in their study, the re-entrance effect takes place in the paranematic phase, and the concentration at which percolation in the nematic is lost does not seem to depend on the connectivity range, at least in the absence of an external field.
		
		Here, we find that percolation is typically not lost within the weakly ordered paranematic phase, but within the nematic phase or across the paranematic-nematic phase transition.
		The re-entrance concentration in the absence of an external field strongly depends on the value of the connectivity range $\lambda/D$.
		These discrepancies between our predictions and those of Ref.~\cite{Otten2012} presumably arise due to their analytical expansion of the contact volume in Legendre-polynomials, which might be inaccurate for large field strengths or particle concentrations.
		
		Investigating how our predictions change for more realistic particles of lower aspect ratios than infinity, we find that an additional high-density percolation threshold arises next to the contained percolation island, which does not exist in the infinitely slender rod limit.~\cite{Finner2019PRE}
		Also, the range of field strengths $K$ in which re-entrance takes place becomes much wider than for infinitely slender particles.
		This indicates that our observed re-entrance effect might actually be observable in real experimental systems, in which also deviations from our model like kinks, particle flexibility and polydispersity come into play.

		The correlation lengths along and perpendicular to the field direction, which are a measure for the physical cluster dimensions below the percolation threshold, confirm earlier predictions that nanoparticle clusters are strongly elongated in the nematic phase, and weakly elongated in the paranematic.~\cite{Otten2012, Finner2019PRL}
		While the bulk percolation threshold in the thermodynamic limit is the same in both directions, this indicates that, in a finite-sized system like a thin film, both steric interactions and external fields may be used to produce materials with anisotropic transport properties.~\cite{Wang2008, Gong2015}
		
		It remains an open question whether in other symmetry-broken phases like the smectic or the columnar phase percolation may be achieved in one or two directions, while keeping the material an insulator in the remaining dimensions.
		Also, at this point in time, no theoretical framework is available to investigate dynamic percolation in flow fields that do not allow for a quasi-static treatment, like shear flow.
		This, we are currently investigating in our ongoing work.\\

%%%%%%%%%%%%%%%%%%%%%%%%%%%%%%%%%%%%%%%%%%%%%%%%%%%%%%
	\begin{acknowledgments}
		S.\,P.\,F.\ and P.\,v.\,d.\,S.\ are funded by the European Union's Horizon 2020 research and innovation programme under the Marie Sk\l{}odowska-Curie grant agreement No 641839.
	\end{acknowledgments}
		
	\bibliography{lit}

\end{document}